# Analysis of Errors - A Support System for Teachers to Analyze the Error Occurring to a Novice Programmer

[1] Aniket Bhawkar, [2] Rohit Belsare, [3] Fenil Gandhi, [4] Pratiksha Somani

[1, 2, 3, 4] Department of Information Technology, VIIT, Pune, Maharashtra 411048, India

**Abstract**

For a novice programmer, coding is equivalent to a nightmare. A novice programmer tries to replicate steps provided by the faculty and on compilation gets a number of errors which the novice programmer is not able to resolve. This system provides support to the faculty about the coding ability of the students and their ability to solve those errors. Also, the faculty can provide a solution to the errors which are occurring to the students and the solution is displayed accordingly. The emphasis of this paper is on developing this system within JAVA and making use of Online Compilers. Moreover, we focus on a new system which is able to provide online code management and these codes get compiled using an online compiler and these programs can be viewed by the respective faculty for cross verification. This paper takes into account the syntactic errors, runtime and semantic errors.

## 1. Introduction

With the increase in demands of Information Technology market, the need of proper trained Computer Professionals is increasing rapidly. This is indirectly increasing the pressure on the schools and engineering institutes to get more and more Computer students trained. That is the main reason for computer programming education to be crucial in the modern era. A novice computer programmer often tries to replicate the tutorials those which are taught in the lectures, practice similar programs and finally try coding them. In this process, the novice programmer gets numerous errors which he/she is not able to deal with. At the same time, a faculty has a difficult task to teach a bunch of students in a limited time.

Additionally, the teacher has to know in Real-Time the amount of concept grasped by the student and their ability to solve the errors which the get as an output after compilation or during Run-Time. Several students often get compilation errors, like syntax-error because of miss typing. Therefore they feel that programming is too hard, which indirectly reduces their moral confidence. The most important problem is the student can't make effective use of the error messages from the compiler. These messages are difficult to understand for novice programmers because the message only indicates a syntactic error and where it occurred. To solve the problems, we needed a system that could present the teacher with an information about the programming capability of the student in real time and their ability to solve the errors which the get during this process. Moreover, the previous research papers failed in providing features like Real-Time Solution Presentation, analysis of Run-Time Errors, and Code-Management for further reference. The proposed paper is written to deal with a system to analyze the errors. Also we provide a system architecture which provides a support to the faculty with the help of which he can teach each and every programming language.

## 2. Literature Survey

This section includes the work already done on this system by various researchers using different methodologies and algorithms. Following is the brief description of some of them:

### 2.1 A Support System for Teaching Computer Programming Based on the Analysis of Compilation Errors

This paper is proposed by Yasuhiko Morimoto, Kunimi Kurusawa, Setsuo Yokoyama, Maomi Ueno, Youzou Miyadera [1]. This paper gives a detailed information about the errors which the students get, are displayed to the respective faculty in an analyzed format. But there was no such provision to provide a real-time solution to the students for the respective errors. This paper only deals with the compilation errors, but was not supposed to deal with the run-time errors.

### 2.2 A Concept of Agent based learning system for C Programming

This paper is proposed by Kazuhiko Nagao and Naohiri Ishii [2]. This paper deals with the analysis of C Programs

and providing the error messages in a simplified format which is easy for novice students to understand. This system is proposed only for UNIX system. Also this system is not able to provide errors in real-time to the teacher. Moreover it also fails in providing solutions to the errors in real time to the students. It fails to capture semantic errors and runtime errors like segmentation fault.

### 2.3 Classroom Experience with Jeroo

This paper is proposed by Dean Sanders and Brain Dorn [3]. This system is useful for learning basic concepts of using object to solve problem and also to write methods that define the behaviour of the object. This paper is useful for novice programmers. In this paper, the syntax is close to Java and C++ and moreover it provides an animated execution of the code and code highlighting. The features of Jeroo are that we can perform file manipulation like new, open, save, etc.

### 2.4 Analysis of Student Programming Errors in Java Programming Courses

This paper is proposed by Ioana Tuugalei Chan Mow [4]. This system provides a proper analysis of several errors which occur while student is programming a JAVA code. In this system, the errors are categorized and the frequency of the error occurrence is calculated. The system provided support for Programming errors like syntactic errors, semantic errors and logical errors. This system failed to provide code management. In this system, the support was only provided for Java Program developers; however the system failed to provide a support system for more than one programming language; the same concept can be implemented for novice programmers who fear to perform programming in various languages.

### 2.5 Observed Failures in Previous Systems

- Capture Run-Time Errors
- Present Real-Time solutions to novice programmer
- Retrieve previous recommended solutions.
- Provide Code Management
- Provide System Architecture which supports every programming language.

## 3. Proposed Methodology

This system is proposed to support the students as well as present the teacher with the errors which are occurring to the students. A database is needed to store all the programs which are coded by the students. Also there is a need of a database which stores all the errors which a student gets. Moreover, there is a need of a database which stores all the solutions provided by the teacher and a database which helps the teacher by providing cause of error. Hence there is a need of four databases and seven segments.

The databases required are :

1. Code Management Database
2. Error Database
3. Solution Database
4. Error Reason Database

The segments required are :

1. Program Gathering Segment
2. Error Gathering Segment
3. Solution Gathering Segment
4. Solution Presentation Segment
5. Programming State Segment
6. Error Reason Extrapolation Segment
7. Error Trend Presentation Segment

### 3.1.1 Code Management Database

Code Management is one of the crucial parts of the system. This Database allows the student to have a check of the various codes that he has performed. Moreover it also allows the teacher to cross-verify the programs those which are coded by the students. This Database is used with the help of Program Gathering segment and the Programming State Presentation Segment. Student can manipulate all those programs which are coded by him, by using the following functionality viz., New, Open, Save, Save As.

### 3.1.2 Error Database

The errors from the Error Gathering Segment are stored in the Error Database. The errors of each and every student are stored in this Database. The respective faculty can view the errors accordingly using this Database.

### 3.1.3 Solution Database

This Database is used to store all the possible solutions of a particular error. The Solution Database is used to store solution which is provided by the faculty. This solution is then provided to the student. On the next occurrence of a similar error, the Solution Database provides the respective solution to the student in Real-Time.

### 3.1.4 Error Reason Database

Error Reason Database stores all possible errors which can occur while coding and also it contains the reason of that error. Most of the compilation errors occur due to miss-typing. This Database provides a backbone to system of overcoming the error.

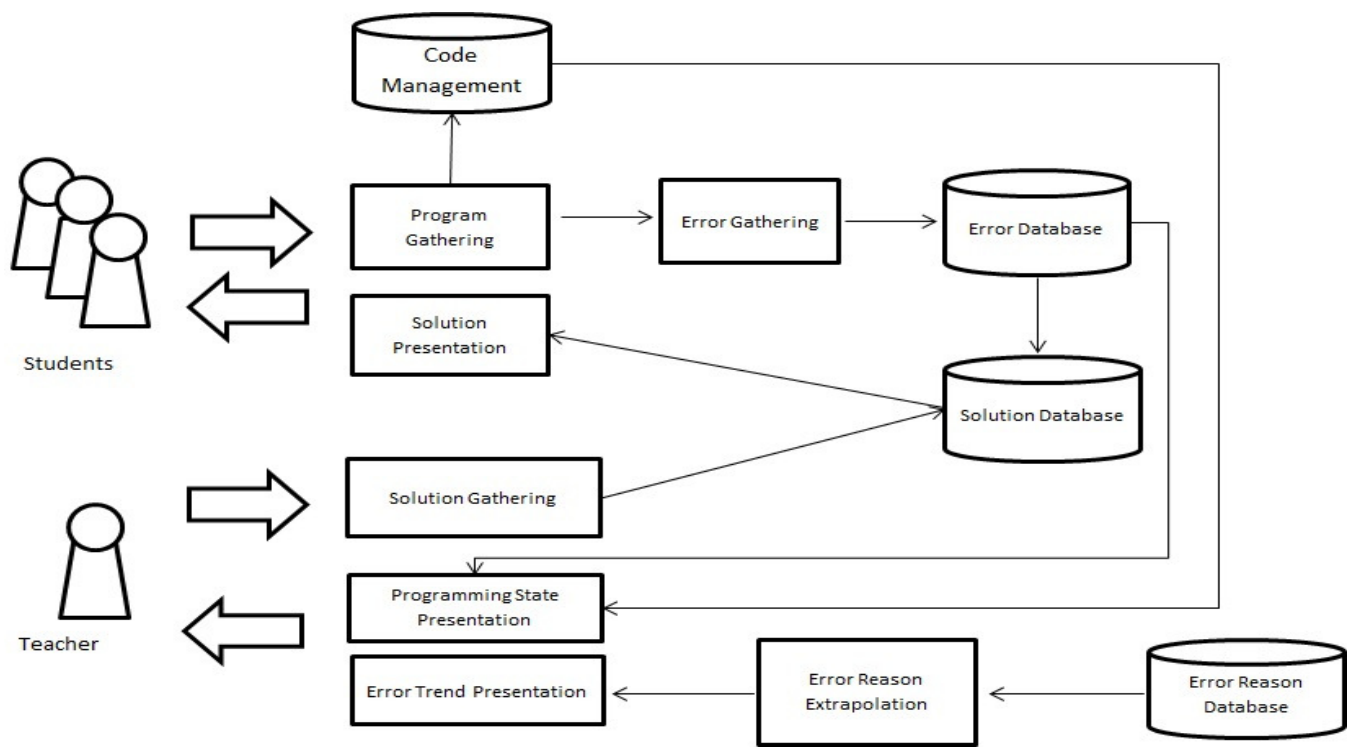

Figure 1. Proposed Analysis of Errors System Architecture

### 3.1.5 Program Gathering Segment

This segment is the initial state of the system. The Program Gathering segment is used to interact with the student. The programs that are coded and compiled by the students get fetched by the Program Gathering Segment. This code is then stored in this Code Management Database. Also the errors that arise in the compile time or run-time are passed on to the error collection segment.

### 3.1.6 Error Gathering Segment

This is the most vital segment of the system. The program which is acquired from the Program Gathering Segment gets compiled by an online compiler and the errors which come as output are collected in the Error Gathering Segment. Errors may not only refer to compilation errors but also run-time errors like Null Pointer, Segmentation Fault etc.

### 3.1.7 Solution Gathering Segment

The faculty can provide a solution to a particular error, like a sample code, or a statement or a precaution. This solution gets mapped to that error in the Solution Database. Whenever a student encounters with a similar type of error, the respective solution can be presented. The Solution Gathering Segment fetches the solution which is provided by the faculty and maps it to the system.

### 3.1.8 Solution Presentation Segment

The solution which is provided by the faculty to a particular error is saved in the Solution Database. Whenever a student arises with an error which has a solution present in the Solution Database, the Solution Presentation Segment presents the student with the solution of that error in real-time. This feature provides the student with a better understanding of the error and also increases his confidence regarding programming.

### 3.1.9 Programming State Presentation Segment

This segment determines the programming situation of the student. The faculty gets an overview about all the efforts which are taken by the student to perform programming. The Programming State Presentation segment is linked to the Code Management Database and the Error Database. Hence the faculty can view all the programs which are coded by all students or a particular student to be specific. It also provides the faculty with the list of errors which were encountered by the students and their efforts taken to solve them. It also provides with a report about the number of times, a student compiled the code to get an appropriate result without error. This segment having being connected to the Code Management Database provides the faculty the exact programs which are coded by the novice programmer and whether they are properly working or not. It also gives a summary to the faculty whether it is required to reteach a topic to the students to clear their misconception about a particular concept.

### 3.1.10 Error Reason Extrapolation Segment

The Error Reason Extrapolation Segment is interlinked with the Error Reason Database and the Error Trend Presentation Segment. This segment provides the faculty with the reason of errors each and every student is getting and provides a report to the faculty about the severity of that error.

### 3.1.11 Error Trend Presentation Segment

This segment presents a list of errors which were encountered by the students and the reason of that error getting raised in the code. It is linked with the Error Reason Extrapolation segment and thus presents a report to the faculty about the errors and their causes. It also helps the faculty in providing a common solution or a hint to the errors.

## 4. Conclusion

The study of the proposed system done so far provides a brief description to determine the ways in which we can help in overcoming the programming difficulties of a novice programmer. The system uses JAVA and an online compiler, is a client-server application, and is not dependent on the barrier Operating Systems. The proposed system overcomes the restrictions of Real-Time Error Presentation to the Faculty and providing a Real-Time Solution to the novice programmer who gets a particular error. Also the system overcomes the drawbacks from previous papers like capturing Run-Time Errors, Code Management, Cross-Verification of the coded programs by the programmer to the respective faculty.

**Acknowledgments**

An endeavor is successful only when it is carried out under proper guidance & blessings. We would like to thank people who helped us in carrying out this work by lending invaluable assistance to us in carrying out this work. We are hereby thankful to Prof. Mr S.R. Sakhare, Head of Department, Information Technology, VIIT, Pune & Prof. Mr Narendra Pathak, Project Coordinator, Department of Information Technology, VIIT, Pune who encouraged at this venture. We sincerely thank Prof. Mrs J.V. Bagade, Project Guide, VIIT, Pune for their constructive & encouraging suggestions. We also thank all Teaching and Non-teaching staff of Department of Information Technology, VIIT, Pune for their kind of co-operation during our course. Finally we are extremely thankful to our Family & Friends who helped us in our work & made the project a successful one.